\documentclass{article}

\usepackage{graphicx}
\DeclareGraphicsExtensions{.jpg,.pdf}
\usepackage{amssymb,amsmath}
\usepackage[english]{babel}
\usepackage{color}

\begin{document}

\title{Biofilms as poroelastic materials}
\author{Ana Carpio, Universidad Complutense de Madrid, Spain \\
Elena Cebri\'an, Universidad de Burgos, Spain \\
Perfecto Vidal, Universidad Complutense de Madrid, Spain }

\date{Aug 17, 2018}

\maketitle

{\bf Abstract.} Biofilms are bacterial aggregates encased in a self-produced
polymeric matrix which attach to moist surfaces and are extremely resistant to 
chemicals and antibiotics. Recent experiments show that their structure is
defined by the interplay of elastic deformations and liquid transport within
the biofilm, in response to the cellular activity and the interaction with the 
surrounding environment.
We propose a poroelastic model for elastic deformation and liquid transport 
in three dimensional biofilms spreading on agar surfaces. The motion of the
boundaries can be described by the combined use of Von K\'arm\'an type 
approximations for the agar/biofilm interface and thin film approximations
for the biofilm/air interface. Bacterial activity informs the macroscopic 
continuous model through source terms and residual stresses, either
phenomenological or derived from microscopic models. We present a 
procedure  to  estimate  the structure of such residual stresses, 
based on a simple cellular automata description of bacterial activity. 
Inspired by image processing, we show that a filtering strategy
effectively smooths out the rough tensors provided by the stochastic
cellular automata rules, allowing us to insert them in the macroscopic 
model without numerical instability.


\section{Introduction}
\label{sec:intro}

The evolution of multicellular systems implicates biological, chemical and
physical processes over a variety of spatial and temporal scales. At a 
microscopic level, cells are discrete entities which perform  tasks (growth, 
division, differentiation, secretion of chemicals, motion, death) in response 
to continuous fields (concentrations of oxygen, nutrients and waste, 
flows, stresses). Simultaneously, individual cells aggregate to form clusters 
exhibiting collective behaviors.  Being able to understand and to
reproduce the dynamics of multicellular systems requires the introduction of 
adequate mathematical models, as well as  suitable tools for their
analysis and simulation.

We consider here bacterial biofilms growing on agar/air interfaces. 
Roughly speaking, a biofilm consists of an elastic solid 
matrix (bacterial cells plus self produced polymeric meshwork) with 
inter-connected pores occupied by an extracellular fluid solution 
\cite{flemming,kapellos}.  Biofilms are known to provide mechanisms
for antibiotic resistance \cite{birnir} and constitute a main source of hospital 
acquired infections \cite{hoiby}.  Understanding their structure may help 
to fight them. 

For specific biofilms, such as those produced by {\em Bacilus Subtilis}, recent
experiments suggest  that the spread and shape of the biofilm
is determined by the interplay of inner liquid transport and elastic deformations 
triggered by the cellular activity and by the interaction with the environment.
Once bacteria adhere to a surface, they differentiate in several types in 
response to local variations created by growth, division, nutrient consumption,
waste production, and cell-cell communication \cite{hera}. Some of them secrete 
exopolymeric substances (EPS)  which  form the extracellular 
matrix (ECM). EPS production changes the osmotic pressure within the biofilm, 
driving water from the agar substrate inside the film and fostering its spread on 
the surface \cite{seminara}. Additionally, the polymeric matrix gives the biofilm a 
certain cohesion, allowing for measurements of elastic Young moduli. Localized 
death in regions of high density and high biochemical stress, combined with 
compression caused by division and growth, trigger the onset of wrinkle formation 
\cite{asally}. As the biofilm expands, complex wrinkled patterns develop. The occurrence 
of successive wrinkle branching and wrinkled coronas is related in \cite{espeso} to 
stiffness gradients created by heterogeneous cellular activity and water migration. In 
later stages, the network of wrinkles becomes a network of channels 
which sustain the development of the biofilm transporting water, nutrients and 
waste \cite{wilking}. The possibility of delamination and folding is 
analyzed in \cite{benamar} by means of neo-Hookean models. We consider
here the biofilm-substrate system as a block.
Whereas biofilm spread due to water absorption from agar has 
been explained using two phase flow models and thin film approximations for the 
biofilm/air interface \cite{seminara}, wrinkle formation has been  reproduced by 
means of plate Von K\'arm\'an equations for the biofilm/agar interface  \cite{espeso}. 
Seeking for a unified representation of both types of processes, we may consider 
 poroelastic descriptions. 

Poroelasticity studies the interaction of fluid flow and deformation in a
fluid-saturated porous medium. This theory was proposed by Biot \cite{biot}
in connection with soil consolidation models for the settlement of structures.
It has later been applied to bone \cite{porobone}, tumor \cite{porotumor}
and tissue \cite{porotissue} studies. There are three major approaches
to the establishment of the main basic equations. The {\it effective} approach
stems from solid mechanics. Effective parameters and constitutive
laws are found averaging over representative volume elements
\cite{biot,rice}. Instead, the {\it mixture} approach originates in the fluid 
mechanics tradition. Each position is occupied by particles of the different
constituents of a mixture. The different species are assigned a density,
stress, energy, and so on. Balance laws for the mass, momentum, 
and energy of each constituent are proposed, from which balance laws
for the mixture are derived \cite{kapellos}. The basic equations provided
by the mixture and the effective approach are similar. However,
the mixture framework is advantageous when more than two constituents
are present and in relative motion, see \cite{cartilage} for articular cartilage,
for instance. In general, the mixture approach provides a better insight on
fluid aspects whereas the effective approach allows for a better interpretation
of parameters associated with the solid phase. A third 
{\it homogenization} approach \cite{keller} uses homogenization techniques
to systematically derive macroscopic equations taking into account
the microscopic structure, relating the effective parameters to the structure
of the phases and analyzing wave propagation aspects of the theory.
Depending on the volume fractions of fluid and solid, the dynamic viscosity
of the fluid, the Lam\'e constants for the solid, the density of the solid,
the hydraulic permeability of the fluid/solid system, the characteristic
time for changes in the displacement of the solid, and the characteristic
length of the system in the macroscopic scale, the fluid/solid system
can be considered as monophasic viscoelastic, monophasic 
elastic, or truly biphasic mixture/poroelastic \cite{kapellos,keller,klapper}.
Applications of these models in geophysics \cite{chen,ghassemi}
and biomedicine \cite{porobone,porotumor,porotissue} usually
fix a spatial region and study the evolution of the different phases
and physical magnitudes in it, avoiding to track moving boundaries,
a relevant aspect to understand biofilm spread and shape.

In this paper, we propose a biphasic mixture/poroelastic description of
biofilms spreading on agar surfaces which takes into account osmotic
flow and incorporates nonlinear effects in the vertical displacements. 
This description should allow us to study simultaneously biofilm
spread due to water intake from agar and the formation of wrinkles 
due to stiffness gradients. The motion of the two interfaces, the interface 
biofilm/agar and the interface biofilm/air is essential in these studies.
We suggest effective  equations for the dynamics of both interfaces.

The paper is organized as follows.  Section \ref{sec:poroelastic}
describes the equations for the deformation of the solid biomass and for
the fluid flow, as well as the balance laws for the biomass and fluid phases
and for the dissolved chemicals.  Sections \ref{sec:vonKarman}
and \ref{sec:lubrication} propose reduced equations for the dynamics
of the biofilm/agar and biofilm/air interfaces, respectively.
Section \ref{sec:averaging} explains how to connect this model
to stochastic discrete representations of cellular activity, such as cellular 
automata models. A mathematical procedure to generate smooth
residual stresses from the cellular automata evolution which should
then be plugged into the poroelastic equations is presented.
Finally, Section \ref{sec:conclusions} summarizes our conclusions.

\section{Poroelastic description of a biofilm spreading on an agar/air interface}
\label{sec:poroelastic}

\begin{figure}[h]
\centering
\hskip -3.5cm (a) \hskip 5cm (b) \\
\includegraphics[width=11cm,angle=0]{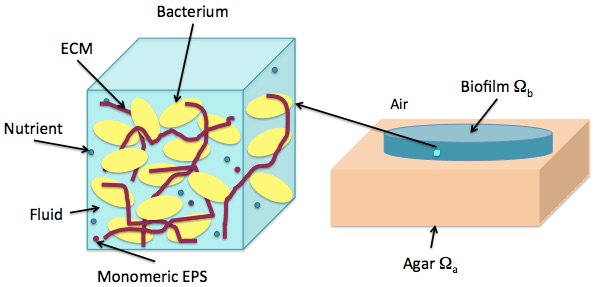}  
\caption{Schematic representation of a biofilm: (a) Microscopic structure
containing biomass (cells, polymeric mesh forming the extracellular matrix), 
fluid and dissolved substances,
(b) Macroscopic view of the configuration under study: a biofilm growing
on an agar block in contact with air.}
\label{fig1}
\end{figure}

In this section, we propose a set of basic equations governing a biofilm in 
expansion on an air/agar interface. The biofilm occupies a region $\Omega_b(t)$ 
over an agar block $\Omega_a(t)$, see Fig. \ref{fig1}. In principle, the biofilm
is formed by bacterial cells, EPS matrix and interstitial fluid. {\it Bacilus Subtilis}  
cells differentiate into different kinds in a biofilm
\cite{hera} (normal cells, surfactin 
producers, EPS producers, inert cells) and may also die. Moreover, the
fraction of EPS matrix in the mixture depends on the type of bacteria.
In biofilms  formed by {\it Pseudomonas} strains spreading in flows, bacteria 
are  scattered in large fractions of extracellular material \cite{stonepnas}.
Instead,  in the biofilms formed by {\it Bacilus Subtilis}  
considered here, bacteria are densely packed,
glued together by small fractions of extracellular material 
\cite{seminara,asally}. This suggests considering a volume
fraction of biomass $\phi_b$ that includes the volume
fraction of cells and ECM (extracellular matrix):
\[
\phi_b(\mathbf x,t) = \phi_{bacteria}(\mathbf x,t) + \phi_{ECM}(\mathbf x,t).
\]
Assuming that no voids neither air bubbles form inside the biofilm, the
mixture is fully saturated, and the volume fraction of fluid $\phi_{\ell}$ 
is given  by  
\begin{equation}
\phi_{\ell}=1- \phi_b.
\label{saturation}
\end{equation}
The values of densities measured for tissues  and
agar do not differ much from the density of water $\rho_w = 10^3 \,
{\rm kg}/{\rm m}^3$ (relative differences of order $10^{-2}$).
Therefore, we will also take the densities of all components to be
constant and equal to that of water \cite{seminara}: $\rho= \rho_w.$
Fluid flow in the biofilms under consideration is a combination of
Darcy and osmotic flow \cite{seminara}.
The equations for a poroelastic material in which water flow is a 
combination of Darcy and osmotic flow, caused by diffusion of 
a certain chemical, are given in  \cite{chen,ghassemi}. We 
revise them next in a more general context, incorporating biomass 
production due to nutrient consumption.

\subsection{Elastic deformations}
\label{sec:elastic}
 
Inspired by \cite{barbour} and \cite{ghassemi},  Ref. \cite{chen} proposes a 
constitutive stress-chemical concentration-strain 
relationship between the volumetric strain increment $d \varepsilon_{V}$,
the effective mean stress increment  $d (\sigma + p) $  and the osmotic
pressure increment $d\pi$:
\begin{eqnarray*}
d   \varepsilon_{V} =  m_V d (  \sigma + p) - m_{\pi} d\pi, 
\end{eqnarray*}
where $\sigma$ is total mean stress (positive in tension), $p$ is the excess pore 
fluid pressure (positive in compression) and $\pi$ the osmotic pressure
(positive in compression). The osmotic pressure created by a chemical
concentration in solution $c_e$ is approximated by the Van't Hoff equation:
\begin{eqnarray}
\pi = {R T \over M_s} c_e = \beta c_e, \label{osmotic}
\end{eqnarray}
where $R$ is a gas constant, $T$ the absolute temperature, 
and $M_s$ the molar mass of the solute. 
The resulting stress-strain relation for an isotropic medium
is \cite{chen,ghassemi}:
\begin{eqnarray}
\sigma_{ij}=  \lambda \delta_{ij} \varepsilon_{kk}({\mathbf u}) + 
2 \mu \varepsilon_{ij}({\mathbf u})
+ m_\pi'  \, c_e \delta_{ij} - \alpha \, p \delta_{ij}, 
\label{constitutiveisotropic}
\end{eqnarray}
where ${\mathbf u}$  is the displacement vector of the biomass
and $ \varepsilon({\mathbf u})$ the deformation tensor:
\begin{eqnarray}
\varepsilon_{ij}({\mathbf u})= {1\over 2} \left(  {\partial u_i \over \partial x_j } 
+ {\partial u_j \over \partial x_i}  \right) + \varepsilon_{ij}^0.
\label{deformation}
\end{eqnarray}
The parameters $\lambda$ and $\mu$ are the Lam\'e constants
of the biomass, related to the Young $E$ and Poisson $\nu$ moduli by
\begin{eqnarray}
\lambda = {E \nu \over (1+\nu) (1-2\nu)}, \quad
\mu = {E \over 2 (1+\nu)}.
\label{moduli} 
\end{eqnarray}
The values of $\alpha$ and $m_\pi'$ depend on assumptions made about
incompressibility and properties of the constituents \cite{ghassemi,chen}.
When the fluid/solid system is incompressible \cite{kapellos}, $\alpha=1$. 
The tensor $\varepsilon_{ij}^0$ represents residual strains in the biomass 
created by growth, swelling or other processes \cite{espeso}.

The motion of the system should then be governed by the evolution equations
\begin{eqnarray}
\rho {\mathbf u}_{tt} - {\rm div(\boldsymbol \sigma({\mathbf u}))} = \rho {\mathbf f},
\label{eq:sigmaev}
\end{eqnarray}
$\rho$ being the density of the medium and $\rho {\mathbf f}$ the body force.
We fix the displacements at the agar/biofilm interface, while imposing
no stress boundary conditions at the air/biofilm interface.  Inertial terms are
often neglected for tissues.

We can study the whole biofilm-agar system applying equation 
(\ref{eq:sigmaev}) and the constitutive law (\ref{constitutiveisotropic})
to both domains $\Omega_a(t) \cup \Omega_b(t)$. Then, we must solve 
(\ref{eq:sigmaev}) with coefficients varying sharply (possibly discontinuous) 
at the agar/biofilm interface, imposing transmission conditions at the interface and fixed
displacements on the agar boundary. This is a standard situation in elasticity that 
can be handled numerically by finite elements, for instance, but it is quite costly
and poses technical difficulties related to the presence of moving interfaces 
and contact fronts. Similar remarks hold for the equations formulated next in 
Sections \ref{sec:flow}-\ref{sec:chemicals}. 
Within this framework, we could also consider biofilm delamination from agar as done 
for composite materials \cite{delamination} if needed. Here we choose to focus on situations 
in which the biofilm remains attached to the agar substratum to cut the computational 
complexity by using effective equations for the moving biofilm boundaries.

\subsection{Fluid flow}
\label{sec:flow}

The fluid flow through pores is the combination of Darcy flow driven
by standard pressure:
\begin{eqnarray}
{ \mathbf J}_d = - k_h  \nabla p, \label{pressuredriven}
\end{eqnarray} 
where $k_h$ is the hydraulic permeability  
($k_h={k\over \mu_\ell}$, $k$ permeability of the solid,
$\mu_\ell$ fluid viscosity),
and osmotic flow driven by osmotic pressure:
\begin{eqnarray}
{\mathbf J}_\pi =  \beta \omega k_h \nabla c_e, \label{osmosisdriven}
\end{eqnarray}
where $\omega$ is the osmotic efficiency. The total flow is \cite{chen}:
\begin{eqnarray}
{\mathbf J}_\ell = {\mathbf J}_d + {\mathbf J}_\pi  = - k_h  \nabla p 
+ \ \beta \omega k_h \nabla c_e.
\label{totalflow}
\end{eqnarray}

The previous effective equations for the fluid flow and the biomass deformation 
can be related to the momentum balance  for the constituents of the mixture
\begin{eqnarray}
\rho_\gamma \phi_\gamma {\partial \mathbf v_\gamma \over \partial t} =
{\rm div} (\boldsymbol \sigma_\gamma) + \sum_{\chi \neq \gamma}
\mathbf F_{\chi \rightarrow \gamma} 
+ \rho_\gamma \phi_\gamma \mathbf f_\gamma, 
\quad \gamma = b,\ell,
\label{lcmomentum}
\end{eqnarray}
using the following constitutive equations for the stress tensors \cite{kapellos,lanir}
\begin{eqnarray}
\begin{array}{l}
\boldsymbol \sigma_b =  - \phi_ b \,  p \, \mathbf I + 
\lambda_b  \, {\rm div}(\mathbf u_b) \, \mathbf I +
\mu_b (\nabla \mathbf u_b + (\nabla \mathbf u_b)^t), \\[1ex]
\boldsymbol \sigma_\ell =  - \phi_ \ell \, p \, 
\mathbf I - {2\over 3} \mu_\ell  \, {\rm div}(\mathbf v_\ell) \, \mathbf I +
\mu_\ell (\nabla \mathbf v_\ell + (\nabla \mathbf v_\ell)^t),
\end{array}
\label{stresses}
\end{eqnarray}
and a constitutive equation for the interaction force \cite{kapellos,lanir}
\begin{eqnarray}
\mathbf F_{\ell \rightarrow b} = - \mathbf F_{b \rightarrow \ell}
= K (\mathbf v_\ell - \mathbf v_b) + p \nabla \phi_b + \phi_b \nabla \pi_b.
\label{interaction}
\end{eqnarray}
The saturation condition $\phi_b + \phi_\ell =1$ implies
$\nabla \phi_b = - \nabla \phi_\ell,$ expected to be small here.
The concentration forces satisfy $\phi_b \nabla \pi_b=-
\phi_\ell \nabla \pi_\ell,$ see \cite{lanir}.
The velocities are related to the displacements by 
${\partial \mathbf u_\gamma \over \partial t} = \mathbf v_\gamma$,
$\gamma=b,\ell.$ In the presence of gravity, the body forces
$\rho_\gamma \mathbf f_\gamma$ are equal to the gravity force 
$-\rho_\gamma \mathbf g$.

Assuming incompressibility of both phases, and
disregarding the inertial terms (local accelerations)
and the viscous contributions in the fluid stress tensor, the
momentum balance laws yield:
\begin{eqnarray}
\begin{array}{l}
0 =  - \phi_ b \nabla (p - \pi_b)  + \mu_b \Delta \mathbf u_b +
(\mu_b + \lambda_b) \nabla ({\rm div}(\mathbf u_b))
+ K (\mathbf v_\ell - \mathbf v_b), \\ [1ex]
0 =  - \phi_ \ell \nabla (p- \pi_\ell)  - K (\mathbf v_\ell - \mathbf v_b).
\end{array}
\label{lcmomentum2}
\end{eqnarray}
The second equation provides a law for fluid flow:
\begin{eqnarray}
\mathbf v_\ell = \mathbf v_b  + {\phi_\ell \over K} \nabla (-p + \pi_\ell)
= \mathbf v_b  + {\phi_\ell \over K} \nabla (- p +  \beta \omega c_e),
\label{lcmomentum3}
\end{eqnarray}
which is related to (\ref{totalflow}) setting $\pi_\ell= \beta \omega c_e$
\cite{osmoticflow} and $K ={\phi_\ell^2 \over k_h}.$
On the other hand, adding up the equations (\ref{lcmomentum2}) we find effective
equations for the displacements \cite{lanir}:
\begin{eqnarray}
\begin{array}{l}
0 =  -  \nabla p + \mu_b \Delta \mathbf u_b +
(\mu_b + \lambda_b) \nabla ({\rm div}(\mathbf u_b)).
\end{array}
\label{lcmomentum4}
\end{eqnarray}

\subsection{Mass balance for biomass and fluid fractions}
\label{sec:mass}
 
Assuming constant and equal density $\rho$ for all the components, the balance laws for the fractions 
of biomass $\phi_b$ and liquid  $\phi_\ell = 1- \phi_b$ are given by
\cite{chen, seminara}:
\begin{eqnarray}
{\partial \phi_b \over \partial t} + {\rm div}  (\phi_b {\mathbf v}_b) = 
 r_b(\phi_b,c_n), 
\label{lcsolid} \\
{\partial \phi_\ell \over \partial t} + {\rm div} (\phi_\ell {\mathbf v}_b + {\mathbf J}_\ell) = - r_b(\phi_b,c_n),
\label{lcfluid}
\end{eqnarray}
where  $r_b(\phi_b,c_n)$ represents  biomass production due to
nutrient  consumption and
${\mathbf v}_b={\partial {\mathbf u}_b \over \partial t}$ is the biomass velocity. 
The displacements ${\mathbf u}_b$ are computed from the equations for the stress
in Section \ref{sec:elastic}. The velocity of the fluid is
\begin{eqnarray}
{\mathbf v}_\ell = {\mathbf v}_b + {{\mathbf J}_\ell \over \phi_\ell}.
\label{velfluid}
\end{eqnarray}
Equation (\ref{lcsolid}) is supplemented with zero boundary conditions when
$\mathbf v_b \cdot \mathbf n <0$, $\mathbf n$ being the outer normal vector. 
Equation (\ref{lcfluid}) requires boundary conditions when $\mathbf v_\ell \cdot \mathbf n <0$.
We set the fluid volume fraction equal to zero when this happens at the interface with air,
and equal to a given value at the interface with agar. To be more realistic, (\ref{lcfluid})
should be coupled to a similar system for water dynamics in agar \cite{seminara} using
transmission conditions at the biofilm/agar interface and given volume fractions at the agar
border.

A simple expression for biomass production is the Monod law \cite{seminara}:
\begin{eqnarray}
r_b(\phi_b,c_n) =  k_b {c_n \over c_n + K_n} \phi_b
\sim {1+\alpha_m \over \tau} {c_n \over c_n + K_n} \phi_b,
\label{sourcebiomass} 
\end{eqnarray}
where $c_n$ is the nutrient concentration, $K_n$ the corresponding
half-saturation constant and
$k_b$ is the production rate, approximated by ${1+\alpha_m \over \tau}$,
$\tau$ being the doubling time for bacteria and $\alpha_m$ a correction
representing EPS production. 

Adding up both conservation laws we find a balance equation for the
whole growing mixture:
\begin{eqnarray}
{\rm div}  (\phi_b {\mathbf v}_b+ \phi_\ell {\mathbf v}_\ell) = 
{\rm div}  (\mathbf v) = 0,
\label{balancemixture}
\end{eqnarray}
where $\mathbf v= \phi_b {\mathbf v}_b+ \phi_\ell {\mathbf v}_\ell$ is the
composite velocity of the mixture. The relative velocity is then
\begin{eqnarray}
\mathbf w= \mathbf v_\ell - \mathbf v_b =
{{\mathbf J}_\ell \over \phi_\ell} =
- {k_h \over \phi_\ell}  \nabla p 
+  \beta \omega {k_h \over \phi_\ell} \nabla c_e,
\label{velrelative}
\end{eqnarray}
where the permeability is often taken to be of the form \cite{porotissue,seminara}
\begin{eqnarray}
k_h(\phi_\ell) = {\phi_\ell^2 \over \zeta}, \quad
\zeta=  {3 \mu_\ell \over 2 R_{bacteria}^2} \phi_b.
\label{permeability}
\end{eqnarray}
Here, $\zeta$ is a friction parameter, $\mu_\ell$ the fluid viscosity and
$R_{bacteria}$ a representative bacterial radius. This
expression for $\zeta$ follows from Stokes theory of viscous drag
applied to the biomass mixture, see \cite{kapellos}
for other choices. The resulting balance equations for both phases
are similar to those proposed in \cite{seminara}, except for the fact that 
the osmotic pressure created by the concentration $c_e$ enters the 
relative velocity $\mathbf w$ and  we have an equation for
$\mathbf u_b$, the velocity being computed as
$\mathbf v_b={d \mathbf u_b \over dt}$.

The mass balance equation for the whole mixture (\ref{balancemixture})
can be rewritten as
\begin{eqnarray}
{\partial \over \partial t}({\rm div} \, \mathbf u_b) =
- {\rm div} (\phi_\ell  \mathbf w) = {\rm div} ( k_h   \nabla p)
-  {\rm div} ( \beta \omega k_h \nabla c_e).
\label{balancemixture2}
\end{eqnarray}
When the biomass is not incompressible,  the left hand
side in this equation becomes 
$ {\partial \over \partial t}( \alpha \, {\rm div} \, \mathbf u_b
+ M^{-1} p)$ \cite{kapellos}.

\subsection{Mass balance for chemicals}
\label{sec:chemicals}

Effective continuity equations for chemical concentration in tissues
are presented in \cite{porotissue}. For the limiting concentration
$c_n$:
\begin{eqnarray}
{\partial  c_n \over \partial t} + {\rm div}  (\mathbf v_\ell c_n)  
- {\rm div} (D_n \nabla c_n)   =  r_n(\phi_b,c_n),
\label{lcnutrient} 
\end{eqnarray}
where $D_n$ is an effective diffusivity \cite{effectiveconcentration}.
Denoting by $D_{n,b}$ and $D_{n,\ell}$ the diffusivities in the biomass
and liquid phases $D_n = D_{n,\ell} {3 d - 2 \phi_\ell (d-1) \over
3 + \phi_\ell (d-1)}$, $d=K_{eq} {D_{n,b} \over D_{n,\ell}}.$
The  source $r_n(\phi_b,c_n)$ represents consumption by the biofilm
\begin{eqnarray}
r_n(\phi_b, c_n) \!=\! - \phi_b k_n { c_n \over c_n + K_n},
\label{sourcecn}
\end{eqnarray}
$k_n$ being the uptake rate and $K_n$ the half-saturation constant.
Zero flux boundary conditions are imposed at the air/biofilm interface.
Instead, at the agar/ biofilm interface, we may impose a constant concentration
through a Dirichlet boundary condition. Being more realistic, we will
couple this diffusion equation to another one defined in the agar substratum 
with zero source and transmission conditions at the interface \cite{espeso}.

The polymeric substances secreted by the cells admit different treatments.
In principle, we have a concentration of monomers which interact to
form polymers of increasing length. The dissolved concentration of
monomers $c_e$ may be selected as the concentration driving the 
osmosis process. It obeys an effective equation:
\begin{eqnarray}
{\partial  c_e \over \partial t} + {\rm div}  (\mathbf v_\ell c_e)  
- {\rm div} (D_e \nabla c_e)   = r_e(\phi_b,c_e),
\label{lceps}
\end{eqnarray}
where $D_e$ is an effective diffusivity, as before.
The  source $r_e(\phi_b,c_e)$ represents monomer production by the biofilm
bacteria:
\begin{eqnarray}
r_e(\phi_b, c_e) \!=\!  \phi_b k_e \, {\rm max}\left(0, 1-{c_e \over c_{e,{\rm max}}}\right),
\label{sourcece}
\end{eqnarray}
$k_e$ being the production rate and $c_{e,{\rm max}}$ a maximum cut-off
value. In practice we should substract a term representing the
monomers that become polymers and form the matrix $\phi_{ECM}$
that we have included in the biomass.  
Zero flux boundary conditions are imposed at the  air/biofilm interface.
At the agar/biofilm interface, we may either impose zero flux boundary
conditions, or couple this diffusion equation to another one defined in the agar 
substratum with  zero source and transmission conditions at the interface.

Alive cells generate waste products which may hinder growth, causing
damage and death \cite{asally,birnir}. The evolution of the concentration of waste 
$c_w$ in  the biofilm may be described by
\begin{eqnarray}
{\partial  c_w \over \partial t} + {\rm div}  (\mathbf v_\ell c_w)  
- {\rm div} (D_w \nabla c_w)  = r_w(\phi_b),
\label{lcwaste}
\end{eqnarray}
where $r_w(\phi_b) =   k_w \phi_b $,  $k_w$ being the waste production rate.
Zero flux boundary conditions are imposed at the interfaces with air
and agar. Waste accumulation is only reduced if carried away by 
circulating fluids or if cells die. It may be necessary to distinguish
at least two phases within the biomass $\phi_b=\phi_{alive}+\phi_{dead}.$
Then, the mass balance equation (\ref{lcsolid}) splits in
\begin{eqnarray}
{\partial \phi_\gamma \over \partial t} + {\rm div}  (\phi_\gamma {\mathbf v}_b) 
=  r_\gamma(\phi_{alive},\phi_{dead},c_n,c_w),  \quad \gamma = alive, dead
\label{lcsolid2} 
\end{eqnarray}
with $r_{alive}+r_{dead}=r_b$.
In equations (\ref{lcnutrient}),(\ref{lceps}), $\phi_b$ must be replaced by
$\phi_{alive}$ in the source definitions (\ref{sourcecn}),(\ref{sourcece}).
The new sources are
\begin{eqnarray}
r_{alive}(\phi_{alive},c_n,c_w) =  k_b {c_n \over c_n + K_n} \phi_{alive}
- k_d {c_w \over c_w + K_w} \phi_{alive},
\label{sourcebiomass2} \\
r_{dead}(\phi_{alive},\phi_{dead},c_w) =  
k_d {c_w \over c_w + K_w} \phi_{alive} - k_{abs} \phi_{dead},
\end{eqnarray}
where $k_d$ represents a rate of death and $k_{abs}$ a reabsorption
rate for dead cells.

As mentioned earlier \cite{hera,espeso}, the biomass is in fact composed of 
fractions of extracellular matrix $\phi_{ECM}$, dead bacteria $\phi_{dead}$, 
and alive cells, which include
inert bacteria $\phi_{inert}$, bacteria secreting EPS $\phi_{EPS}$, 
bacteria secreting surfactin $\phi_{surf}$ and bacteria which
do not produce such substances and are able to divide in a normal way
$\phi_{normal}$. In practice, we may have to distinguish the different
species:
\[
\phi_b=\phi_{normal}+\phi_{dead}+\phi_{inert}+\phi_{EPS}+\phi_{surf}+\phi_{ECM},
\]
which would require introducing additional balance equations for the different 
fractions and their interaction.

\section{Motion of the interfaces}
\label{sec:interfaces}

The equations presented in Section \ref{sec:poroelastic} are suitable
to represent elastic deformations and liquid transport in a biofilm. However,
as the biofilm swells and deforms, its boundary moves.
To simplify the study of the evolution of a three dimensional biofilm
it is desirable to obtain reduced equations for the motion
of its boundary. In our geometry, the boundary is formed by two interfaces:
the interface agar/biofilm and the interface air/biofilm. Whereas the first
one controls the formation of wrinkled shapes \cite{espeso}, the second 
one defines biofilm spread \cite{seminara}.

\subsection{Von K\'arm\'an approximation for the agar/biofilm interface}
\label{sec:vonKarman}

Equations for the dynamics of the agar/biofilm interface follow using a 
Von K\'arm\'an type approximation, since the thickness of the biofilms
is small compared to its radius. While  initially flat,
the displacements in the direction orthogonal to the interface may 
become large. Thus, the linear definition of the strain and stress tensors 
in (\ref{deformation}) is replaced by 
\begin{eqnarray}
\varepsilon_{ij}&=&{1\over 2} \left( {\partial u_{i} \over \partial x_{j}} 
+ {\partial u_{j} \over \partial x_{i}}  
+ {\partial \xi \over \partial x_{i}} {\partial \xi \over \partial x_{j}}
\right) + \varepsilon_{ij}^0, \quad i,j=1,2,
\label{strain}\\
\sigma_{11}&=& {E \over 1-\nu^2} (\varepsilon_{11}+ \nu \varepsilon_{22})
+ \sigma_{11}^0, \nonumber \\
\sigma_{12}&=& {E \over 1+\nu} \varepsilon_{12} + \sigma_{12}^0, 
\nonumber \\
\sigma_{22}&=& {E \over 1-\nu^2} (\varepsilon_{22}+ \nu \varepsilon_{11})
+ \sigma_{22}^0,
\nonumber 
\end{eqnarray}
which includes nonlinear terms, as well as residual strains $\varepsilon_{ij}^0$ 
caused by bacterial activity and residual stresses  $\sigma_{ij}^0$ estimated averaging 
the osmotic pressure and fluid pressure contributions to  the three dimensional biofilm 
in the law (\ref{constitutiveisotropic}) in the out of plane direction. 
We denote the in-plane displacements 
by ${\bf u}=(u_1(x_1,x_2,t),u_2(x_1,x_2,t))$ and the out of plane 
displacements of the interface by $\xi(x_1,x_2,t)$. 
The coordinates $(x_1,x_2)$ vary along the 2D 
projection of the  3D biofilm structure on the biofilm/agar interface.
Identifying the biofilm with an elastic film growing on a 
viscoelastic agar substratum, the interface motion is governed by the
equations \cite{huang,espeso}:
\begin{eqnarray}
{\partial \xi \over \partial t} = {1 - 2 \nu_v \over 2 (1-\nu_v)} {h_v \over \eta_v}
\Bigg[ D (- \Delta^2 \xi + \Delta C_M)  
+ h {\partial\over \partial x_{j}} \left( \sigma_{ij}({\bf u}) 
{\partial \xi \over \partial x_{i}}\right) \Bigg]
-{\mu_v \over \eta_v} \xi, \label{plategrowth1bis} \\
{\partial {\bf u} \over \partial t} = {h_v h \over \eta_v} 
{\rm div}({\boldsymbol \sigma({\bf u})}) - {\mu_v \over \eta_v} {\bf u}, 
\label{plategrowth2bis}  
\end{eqnarray}
where $h_v$ is the thickness of the viscoelastic agar substratum and 
$\mu_v$, $\nu_v$, $\eta_v$  its rubbery modulus, Poisson ratio, and  
viscosity, respectively.
The bending stiffness is $D={E h^3\over 12(1-\nu^2)}$,  $h$ being
the initial biofilm thickness. Summation over repeated indexes is intended.
Here, the first equation describes out-of-plane bending 
$\xi$ and the second one in-plane stretching for the displacements 
${\bf u}=(u_1,u_2)$. Modified equations taking into account possible
spatial variations of the parameters are given in \cite{sergei}.

Let us explain now how to estimate the residual strains $\varepsilon^{0}_{ij}.$ 
A growing biofilm is in a state of compression due to cell division and, 
eventually, water absorption, or cell death. In terms of a growth tensor 
$\mathbf g$ \cite{benamar2}, the residual term
$C_M= {\partial (g_{31}+g_{13}) \over \partial x_1} 
+ {\partial (g_{32}+g_{23}) \over \partial x_2}$
and the residual strains $\varepsilon_{ij}^0$ are 
$ \varepsilon_{ij}^0= -{1\over 2} \left( g_{ij} + g_{ji} 
+ g_{3i} g_{3j} \right).$
Plugging residual stresses with this structure into 
(\ref{plategrowth1bis})-(\ref{plategrowth2bis}), we are able to reproduce
wrinkle coarsening and opening up in radial branches observed in
biofilms spreading on surfaces \cite{espeso}.
This phenomenon is associated with the expansion at certain speeds
of compression fronts. 

\begin{figure}
\centering
 (a) \hskip 3cm (b) \\
\includegraphics[height=3cm]{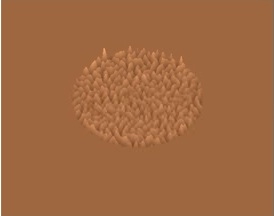}
\includegraphics[height=3cm]{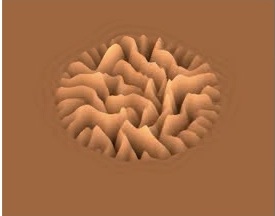} \\
(c) \hskip 3cm (d) \\
\includegraphics[height=3cm]{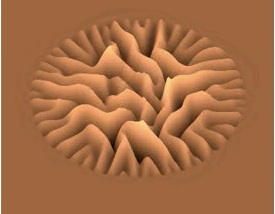}
\includegraphics[height=3cm]{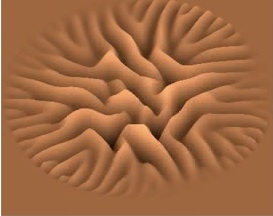}
\caption{Wrinkle  branching and coarsening in a spreading film.}
\label{fig2}
\end{figure}

This approach does not impose a particular shape for the
biofilm, which evolves as dictated by biomass production in response to
variations in the concentration of nutrients, waste and autoinducers.
However, simple illustrative simulations can be performed assuming 
a circular form.
Often $\varepsilon^0_{ij}({\mathbf x},t) = -\varepsilon_0({\mathbf x},t) \delta_{ij}$
with  $ \varepsilon_0>0.$ If we assume that cells do not grow at expense of their 
neighbors, $\delta_{ij}$ is a diagonal unit tensor in polar coordinates in a circular 
film.  Setting $\varepsilon_0({\mathbf x},t) =
\varepsilon_0(|\mathbf x|-ct)$ for adequate velocities $c>0$ and constant or
radially increasing profiles $\varepsilon_0$ we obtain patterns like
those in  Fig. \ref{fig2}. Wrinkled coronas, that is, coronas of radial wrinkles 
issuing from a central core, are associated to the so-called
corona instabilities: a swollen corona with diminished Young modulus 
around a harder core \cite{espeso}.  They can be reproduced
imposing this spatial structure in the Young modulus in the Von K\'arm\'an
equations.

Tracking the biofilm/agar interface by means of equations 
(\ref{plategrowth1bis})-(\ref{plategrowth2bis}) forbids delamination, a phenomenon that has been 
reported in a variety of  thin films coating surfaces \cite{ortiz}. Von K\'arm\'an theory is adapted
in \cite{ortiz} to describe the evolution of films already debonded from a substratum forming
blisters. \cite{napoli} characterizes the onset of delamination in films whose edges are kept 
fixed while growing attached to a substrate. A situation closer to the geometry under study here
is considered in \cite{benamar} by means of a neo-Hookean elastic energy. If one follows branches
of analytical solutions describing circular patches with a moderate adhesion which grow on a stiff 
substrate, contour undulations develop, buckling appears and, ultimately, a regular arrangement
of folds emerges. In our setting, films can develop contour undulations \cite{espeso}, however, we 
do not consider delamination here.

\subsection{Thin film approximation for the air/biofilm interface}
\label{sec:lubrication}

The effective equations for the biofilm deformation in Section \ref{sec:elastic} 
govern $\mathbf u_b$, that is, the displacement of the biomass. However,
the dynamics of the free air/biofilm interface is influenced by the 
displacement of liquids too. We obtain an equation for the motion of the
interface air/biofilm using the mass balance equation for the mixture
(\ref{balancemixture2}). During the first stages of biofilm spread, in which 
the agar/biofilm interface remains flat and the biofilm reaches a height
$x_3=h(x_1,x_2,t)$, we integrate
\begin{eqnarray*}
{\rm div} \,\mathbf v =
{\rm div} \, \mathbf v_b -{\rm div} ( k_h   \nabla p)
+ {\rm div} ( \beta \omega k_h \nabla c_e) = 0,
\end{eqnarray*}
in the vertical direction to get
\begin{eqnarray*}
\int_{0}^h {\partial (\mathbf v \cdot \hat{\mathbf x}_1) \over \partial x_1} \, dx_3
+ \int_{0}^h {\partial (\mathbf v \cdot \hat{\mathbf x}_2) \over \partial x_2} \, dx_3
+ \int_{0}^h {\partial  (\mathbf v \cdot \hat{\mathbf x}_3) \over \partial x_3} \, dx_3 =0,
\end{eqnarray*}
$\hat{\mathbf x}_1$, $\hat{\mathbf x}_2$ and $\hat{\mathbf x}_3$ being the 
unit vectors in the coordinate directions. By Leibniz's rule:
\begin{eqnarray*}
\int_{0}^h {\partial (\mathbf v \cdot \hat{\mathbf x}_i) \over \partial x_i} \, dx_3
= {\partial \over \partial x_i} \left[
\int_{0}^h (\mathbf v \cdot \hat{\mathbf x}_i) \, dx_3 \right]
- \mathbf v \cdot \hat{\mathbf x}_i\big|_h {\partial h \over \partial x_i}, 
\quad i=1,2.
\end{eqnarray*}
Therefore
\begin{eqnarray}
\begin{array}{l}
{\partial \over \partial x_1} \left[
\int_{0}^h (\mathbf v \cdot \hat{\mathbf x}_1) \, dx_3 \right]
+ {\partial \over \partial x_2} \left[
\int_{0}^h (\mathbf v \cdot \hat{\mathbf x}_2) \, dx_3 \right]
\\ [1ex] 
- \mathbf v \cdot \hat{\mathbf x}_1\big|_h {\partial h \over \partial x_1}
- \mathbf v \cdot \hat{\mathbf x}_2\big|_h {\partial h \over \partial x_2}
+ \mathbf v \cdot \hat{\mathbf x}_3\big|_h  =
\mathbf v \cdot \hat{\mathbf x}_3\big|_0.
\end{array} 
\label{height}  
\end{eqnarray}
Notice that $\mathbf v \cdot \hat{\mathbf x}_i = {dx_i \over dt}$, $i=1,2,3$.
Differentiating $x_3(t)=h(x_1(t),x_2(t),t)$ with respect to time we find
\begin{eqnarray*}
\mathbf v \cdot \hat{\mathbf x}_3 \bigg|_h = {d x_3 \over dt} = {d \over dt} h(x_1(t),x_2(t),t) 
&=& {\partial h \over \partial t} + {\partial h \over \partial x_1} {d x_1 \over dt}
+ {\partial h \over \partial x_2} {d x_2 \over dt}  \\[1ex]
&=& {\partial h \over \partial t} + 
\mathbf v \cdot \hat{\mathbf x}_1 \bigg|_h  {\partial h \over \partial x_1} 
+ \mathbf v \cdot \hat{\mathbf x}_2 \bigg|_h {\partial h \over \partial x_2} .
\end{eqnarray*}
Inserting this identity in (\ref{height}) we obtain the equation
\begin{eqnarray}
{\partial h \over \partial t } +
{\partial \over \partial x_1} \left[
\int_{0}^h (\mathbf v \cdot \hat{\mathbf x}_1) \, dx_3 \right]
+  {\partial \over \partial x_2} \left[
\int_{0}^h (\mathbf v \cdot \hat{\mathbf x}_2) \, dx_3 \right]  =
\mathbf v \cdot \hat{\mathbf x}_3\bigg|_0,
\label{height2}
\end{eqnarray}
where
\begin{eqnarray}
\begin{array}{l}
\mathbf v \cdot \hat{\mathbf x}_1 =
{d \mathbf u_{b,1} \over dt} 
- k_h  {\partial p \over \partial x_1} + \beta \omega k_h 
{\partial c_e \over \partial x_1}, \\
\mathbf v \cdot \hat{\mathbf x}_2 =
{d \mathbf u_{b,2} \over dt}  
- k_h  {\partial p \over \partial x_2}  + \beta \omega k_h 
{\partial c_e \over \partial x_2}, \\
\mathbf v \cdot \hat{\mathbf x}_3 =
{d \mathbf u_{b,3} \over dt}   
- k_h  {\partial p \over \partial x_3}  + \beta \omega k_h 
{\partial c_e \over \partial x_3}.
\end{array} \label{height3}
\end{eqnarray}
If the biofilm/agar interface is not flat, $0$ is replaced in the previous
computations by an expression representing the height $\xi$ of this interface,  
calculated in Section \ref{sec:vonKarman}. Additional terms representing the 
spatial variations of $\xi$ would appear in (\ref{height2}). 

When the biomass is considered poroelastic $\mathbf u_b,p,c_e$ 
are governed by the equations in Section \ref{sec:poroelastic}. 
When it can be approximated by a fluid, Stokes equations apply
\cite{seminara} and a closed equation for the height $h$
can be derived by asymptotic techniques assuming
a circular shape.

\section{Coupling to discrete descriptions of cellular activity}
\label{sec:averaging}

In the previous sections cellular activity is represented through 
phenomenological sources and residual stresses. Alternatively,
we may couple the equations for the macroscopic evolution of the
biofilm to a discrete description of the cellular dynamics.  

Cellular automata representations, for instance, furnish a simple 
approach which allows for an easy transfer of microscopic information 
into macroscopic models. We divide the biofilms in cubic tiles. Each of 
them contains a few bacteria. Further simplification of the computational 
geometry identifies each tile with one bacteria.
Then, the matrix is a virtual glue that keeps bacteria together, which
seems reasonable for biofilms on surfaces producing small fractions of
EPS \cite{seminara,picioreanu}.
This strategy will allow us to evaluate growth tensors due to cell activity
(division, death, secretion, water absorption). They are defined on the
cellular automata grid. We can use the same grid to discretize the
equations for macroscopic concentrations and displacement fields.
The residual stresses that enter the F\"oppl-Von K\'arm\'an
equations for the deformation of the agar/biofilm interface can be
computed from the growth tensor provided by the cellular automata
descriptions. Instead, the air/biofilm interface changes now according to
the cellular automata rules to create new tiles representing newborn
bacteria, absorbed water or reabsorbed dead cells and displace the
remaining tiles.
The rules to decide the status of each tile may be based on
probabilities depending on the pertinent concentrations \cite{espeso}
or dynamic energy budget models \cite{birnir}. We sketch next the
first approach. 

\subsection{Air/biofilm interface dynamics and rough residual  strains}
\label{sec:air/biofilm}

The evolution of the air/biofilm interface depends on the creation
of new tiles and the displacement of existing ones.
Assuming the nutrient is the concentration limiting
biofilm growth, we set tiles  ${\cal C}$ occupied by alive bacteria 
to divide with probability \cite{hermanovic}:
\begin{eqnarray*}
P_d({\cal C}) = {c_n({\cal C}) \over c_n({\cal C}) + a_n}, \label{pdiv}
\end{eqnarray*}
with $a_n>0$.  The concentration obeys an equation similar to
(\ref{lcnutrient}) replacing $\phi_b$ with a weight equal to $1$ in
alive cells and $0$ otherwise.
Newborn bacteria push existing cells in  the direction of minimum 
mechanical resistance, that is, the shortest  distance to the air/bioiflm 
interface or to a dead cell.

As an indicator of death due to biochemical stress
we chose the concentration of waste $c_w$.
A tile ${\cal C}$ is scheduled to die with probability:
\begin{eqnarray*}
P_w({\cal C}) = {c_w({\cal C}) \over c_w({\cal C}) + a_w}, \label{pdead}
\end{eqnarray*}
with $a_w>0$.  When surrounded by enough alive cells, dead cells 
may be reabsorbed by the rest, the tile being occupied by a 
newborn cell. Otherwise, a necrotic region is created \cite{birnir}.
The evolution of the concentration of waste $c_w$ in the
biofilm  is governed by equation (\ref{lcwaste}), replacing
$\phi_b$ with a weight equal to $1$ in alive cells and $0$ otherwise. 
A more sophisticated dynamic energy budget treatment of death 
processes can be found in \cite{birnir}.

Taking more autoinducer concentrations into consideration, we
may define additional probabilities for other behaviors, such as
differentiation cascades into autoinducer producers \cite{espeso}.
Probability laws using osmotic pressure would allow to create water
tiles too \cite{espeso}.

\begin{figure}[h!]
\centering
\hskip -5mm (a) \hskip 2.5cm (b) \hskip 2.5cm (c) \\
\includegraphics[width=3cm,height=3cm]{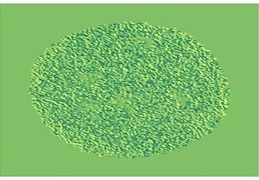}
\includegraphics[width=3cm,height=3cm]{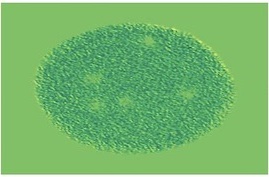}
\includegraphics[width=3cm,height=3cm]{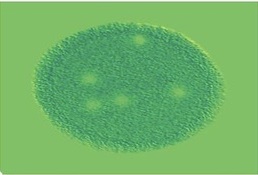} 
\includegraphics[width=0.5cm,height=3cm]{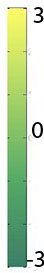}\\
\hskip -5mm (d) \hskip 2.5cm (e) \hskip 2.5cm (f) \\
\includegraphics[width=3cm,height=3cm]{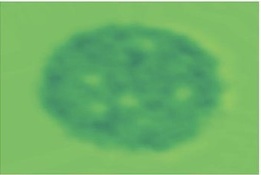}
\includegraphics[width=3cm,height=3cm]{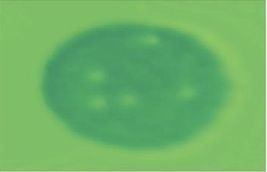}
\includegraphics[width=3cm,height=3cm]{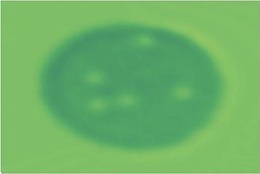}
\includegraphics[width=0.5cm,height=3cm]{fig3g.jpg}
\caption{
(a) $\varepsilon_{11}^0$ component  of the residual strain tensor when $N=1$.
(b) Averaged $\varepsilon_{11}^0$ component when $N=10$.
(c) Averaged $\varepsilon_{11}^0$ component when $N=100$.
(d) Filtered $\varepsilon_{11}^0$ component when $N=1$.
(e) Filtered $\varepsilon_{11}^0$ component when $N=10$.
(f) Filtered $\varepsilon_{11}^0$ component when $N=100$.
The light peaks mark regions where the inicial cell population is larger. In those regions, 
the nutrient concentration depletes while the waste concentration increases, triggering 
cell dead by lack of resources and biochemical stress. Stresses become higher in the outer
ring due to higher availability of resources and higher division rates. 
As the number of averaged $N$ runs of the cellular automata step grows,  
the averages become smoother and  the underlying
spatial structure is better defined. Using filters, the spatial structure
is already visible when $N=1$ and is well defined with a few more trials.}
\label{fig3}
\end{figure}

Let us explain now how to calculate residual strains and stresses.
Keeping track of all the new tiles created and the direction in which their 
predecessors where shifted, we are able to define a growth tensor
$\mathbf g$. To do so, we introduce a vector 
${\mathbf w}=(w_1,w_2,w_3) a,$ where $a$ is the tile size. 
The component $w_1$ is calculated at each 
site by adding $\pm 1$ cumulatively  for each tile shifted in the $x_1$ 
direction in the positive or negative sense, respectively. 
In a similar way, we calculate $w_2$ and $w_3$ along the $x_2$ and $x_3$ 
directions, respectively. Once the resulting vector $\mathbf w$ is normalized to have 
norm $a$, we evaluate $\nabla {\mathbf w}$ approximating  the derivatives 
by finite differences. Finally,  we average all the contributions 
from $\nabla {\mathbf w}(x_1,x_2,x_3)$ over $x_3$ to obtain 
${\mathbf g}(x_1,x_2)$.
However, stochastic variations make this tensor unsuitable to be inserted  
in the  F\"oppl-Von K\'arm\'an equations (\ref{plategrowth1bis})-(\ref{plategrowth2bis}), 
because of numerical instability.

\subsection{Smooth residual strains and agar/biofilm interface  dynamics}
\label{sec:agar/biofilm}

An effective way to smooth out the residual strains is to filter them by image processing 
techniques, which in addition visualizes the underlying spatial structure.
To do so we formulate a denoising problem: given an observed magnitude 
$f^{obs}=f+n$, we seek the primary structure $f$ by removing the noise $n$. 
This problem can be solved applying a split Bregman method to a ROF (Rudin-Osher-Fatemi) 
variational model \cite{torres,osher}:
Find $f$ minimizing $\int |\nabla f| + {\eta \over 2} \int |f-f^{obs}|^2,$ for $\eta>0$
large. The split Bregman method incorporates the constraint $d=\nabla f$, sets
$s(b,f,d)= \int |b + \nabla f -d|^2 $ and implements the iteration:
\begin{eqnarray}
(f^{(k+1)},d^{(k+1)}) = {\rm Argmin}_{(f,d)} \{ |d|
+ {\eta \over 2} \| f^{obs} - f \|_{L_2}^2
+ {\delta \over 2} s(b^{(k)},f,d)
\}, \nonumber \\
b^{(k+1)} = b^{(k)} + \nabla f^{(k+1)} - d^{(k+1)}, \nonumber
\end{eqnarray}
with $\delta>0$. The minimization procedure solves for each variable
alternatively:
\begin{eqnarray}
f^{(k+1)} = {\rm Argmin}_{f} \{ {\eta \over 2} \|f^{obs} - f \|_{L_2}^2
+ {\delta \over 2} s(b^{(k)},f,d^{(k)}) \}, \nonumber \\
d^{(k+1)} = {\rm Argmin}_{d}  \{ |d| 
+ {\delta \over 2} s(b^{(k)},f^{(k+1)},d) \},
 \nonumber \\
b^{(k+1)} = b^{(k)} + \nabla f^{(k+1)} - d^{(k+1)}. \nonumber
\end{eqnarray}
Since the first functional is differentiable,  we  write the Euler-Lagrange 
equation and compute $f^{(k+1)}$ by a Gauss-Seidel  method.
To solve the second optimization problem we use shrinkage operators:
\begin{eqnarray}
d^{(k+1)} = \mbox{\rm shrink} (b^{(k)}+ \nabla f^{(k+1)}, 
{1 \over \delta}),
\nonumber \\
\mbox{\rm shrink}(x,\gamma) = {x \over |x|} 
\mbox{\rm max}(|x|-\gamma,0).
\nonumber
\end{eqnarray}
This strategy leads to the algorithm:
\begin{itemize}
\item Initial guess $f^{(0)} = f^{obs}$, $d^{(0)}=0$, $b^{(0)}=0.$
\item While $\| f^{(k)} - f^{(k-1)} \|_{L_2} > Tol$
\begin{itemize}
\item $f^{(k+1)} = G^{(k)}$, 
\item $d^{(k+1)} = {\rm shrink}(b^{(k)} + \nabla f^{(k+1)}, {1\over
\delta})$,
\item $b^{(k+1)} = b^{(k)} + \nabla f^{(k+1)} - d^{(k+1)}$,
\end{itemize}
\end{itemize}
where, for $j=1, \ldots, M$
\begin{eqnarray}
 G^{(k)}_j = {\delta \over \eta \!+\! 2\delta}
\Big(f_{j+1}^{(k)} + f_{j-1}^{(k)} \!-\! (d_j^{(k)}\!\!-\!d_{j-1}^{(k)}) 
\!+\! (b_j^{(k)}\!\!-\!b_{j-1}^{(k)})\Big)  \!+\! {\eta  \over \eta 
\!+\! 2 \delta} f_j^{obs}, 
\nonumber
\end{eqnarray}
with  $\nabla f^{(k+1)}_j = f^{(k+1)}_{j+1} - f^{(k+1)}_j.$

Fig. \ref{fig3} computes the residual stresses for early stages of the evolution of an 
initially circular biofilm patch containing a few mounds, that is, regions where the biofilm is higher and contains more cells. In early stages, cell division affects mainly the biofilm height, in accordance with \cite{seminara}, though the circular shape may develop irregularities later \cite{espeso,benamar}.
The filtered fields  in Fig. \ref{fig3}(d)-(f) set $f^{obs} = \varepsilon_{11}^{0,av}$  over a 2D grid in the plane $x_1x_2$, relabeling to transform it into a 1D vector. Here, the ensemble averages of the residual strain tensors $\boldsymbol  \varepsilon^{0,i}$ at trials $i=1,...,N$ are denoted by
$\boldsymbol \varepsilon^{0,av}= \sum_{i=1}^N \boldsymbol  \varepsilon^{0,i}$. At each trial we run the cellular automata step, in which new cells are created or killed according the the selected probabilities starting from the same initial configuration in all of them.
For $N$ large enough these ensemble averages allow us to visualize the spatial variations caused by cellular activity,  see Figure \ref{fig3} (a)-(c). 
The resulting average becomes smoother as the number of runs $N$ increases. However, the remaining spikes still cause instability and the computational cost of this process is very high. Instead, this filtering process always produces fields which
are smooth enough to be plugged in (\ref{plategrowth1bis})-(\ref{plategrowth2bis}) without causing numerical instability. Filtered fields
also  reproduce the correct underlying spatial structure for a very low number of runs $N$, lowering drastically the computational cost.

In this framework, the simulations of biofilm behavior would alternate steps in which we update the configuration of biofilm tiles following the cellular automata rules and evaluate the resulting residual stresses, with steps in which the biofilm shape is deformed as determined by the F\"oppl-Von K\'arm\'an equations for the agar/biofilm interface in Section \ref{sec:vonKarman} (see \cite{espeso} for details) and steps in which the poroelastic and concentration equations are solved to update
pressure, displacement, velocity and concentration fields.

\section{Conclusions}
\label{sec:conclusions}

Three dimensional multicellular shapes arise through the interaction of mechanical forces
and cellular activities. Bacterial communities furnish model systems for studying such interplay. 
We consider here biofilms growing on air/agar interfaces, which have
been shown to adopt wrinkled shapes and undergo swelling processes while spreading in 
interaction with the agar substratum. We have proposed a poroelastic solid/fluid model
for the elastic deformation of the biomass matrix and the transport of interstitial
fluid within it. Analyzing the two distinguished interfaces defining the biofilm borders
we have found two different descriptions
for the evolution of each of them. Whereas the interface agar/biofilm is reasonably represented 
by Von K\'arm\'an type approximations, the interface air/biofilm seems to require lubrication type
approaches for thin films. These developments take into account the cellular activity through phenomenological 
sources and residual tensors included in the macroscopic equations.
Instead, we may consider a discrete model of cellular activity to justify such phenomenological 
terms. Coupling to a simple cellular automata representation of bacterial activity, one can define 
residual stress tensors. Image filtering techniques allow us to regularize them  so that can be 
effectively used in macroscopic models without causing instability at low computational cost. 
Whereas the dynamics of the agar/biofilm interface 
can still be described by Von K\'arm\'an type equations containing such residual stresses,
the dynamics of the air/biofilm interface is now determined by the cellular automata rules
for creation and motion of grid tiles.  Delamination effects are not considered
here and would require further study.

\bigskip
{\bf Acknowledgements.} 
This research has been supported by MINECO grants No. MTM2014-56948-C2-1-P
and No. MTM2017-84446-C2-1-R.

\end{document}